\documentclass{emulateapj}

\slugcomment{Submitted to the Astrophysical Journal}

\begin{document}

\title{The Effects of Ram Pressure on the Cold Clouds in the Centers of Galaxy Clusters}

\author{Yuan Li\altaffilmark{1}, Mateusz Ruszkowski\altaffilmark{1}, Grant Tremblay\altaffilmark{2}}

\altaffiltext{1}{Department of Astronomy, University of Michigan, 1085 S University Ave, Ann Arbor, MI 48109; email: yuanlium@umich.edu}
\altaffiltext{2}{Department of Physics and Yale Center for Astronomy \& Astrophysics, Yale University, 217 Prospect Street, New Haven, CT 06511}

\begin{abstract}

We discuss the effect of ram pressure on the cold clouds in the centers of cool-core galaxy clusters, and in particular, how it reduces cloud velocity and sometimes causes an offset between the cold gas and young stars. The velocities of the molecular gas in both observations and our simulations fall in the range of $100-400$ km/s, much lower than expected if they fall from a few tens of kpc ballistically. If the intra-cluster medium (ICM) is at rest, the ram pressure of the ICM only slightly reduces the velocity of the clouds. When we assume that the clouds are actually ``fluffier'' because they are co-moving with a warm-hot layer, the velocity becomes smaller. If we also consider the AGN wind in the cluster center by adding a wind profile measured from the simulation, the clouds are further slowed down at small radii, and the resulting velocities are in general agreement with the observations and simulations. Because ram pressure only affects gas but not stars, it can cause a separation between a filament and young stars that formed in the filament as they move through the ICM together. This separation has been observed in Perseus and also exists in our simulations. We show that the star-filament offset combined with line-of-sight velocity measurements can help determine the true motion of the cold gas, and thus distinguish between inflows and outflows.

\end{abstract}

\keywords{}

\section{Introduction}
Most dynamically relaxed galaxy clusters have a cool core where the temperature of the intra-cluster medium (ICM) is only $1/2-1/3$ of the virial temperature \citep{Fabian1994}. Although these cool-core clusters do not harbor classical cooling flows of hundreds to a thousand solar masses per year, likely due to radio mode AGN feedback \citep{McNamara2007}, many of them develop a ``reduced cooling flow'' \citep{ODea2008}. The detection of H$\alpha$, CO and other emission lines indicate the presence of multi-phase gas in the centers of cool-core clusters \citep{Hu1985, Edge2001}. In fact, the brightest cluster galaxies (BCGs) often contain molecular gas of $10^9-10^{11} \rm M_\odot$, with a typical star formation rate of a few to a few tens of solar masses per year \citep{Hicks2005, Hoffer2012}. 

The cold gas in some clusters is only detected in the nuclei, but in many other clusters, it exhibits clumpy or filamentary morphology with spatial extension of a few to a few tens of kpc \citep{McDonald10}. The clumps/filaments usually have complex dynamics, showing both inflow and outflow, with typical line-of-sight velocities of $100-400$ km/s \citep{Salome2006, McDonald12, Tremblay2016}. In nearby cool-core cluster Perseus, filaments are observed to have typical width below100-500 pc \citep{Conselice2001}. The H$\alpha$ filaments are spatially coincident with soft X-ray features, suggesting that they originate from thermal instabilities of the hot ICM \citep{McCourt12, Sharma2012}. The cold filaments stretch out predominantly radially from the nucleus \citep{Canning2010}. They also coexist with dust lanes \citep{Mittal2012} and sometimes UV features \citep{McDonald2009}, suggesting that some filaments are forming stars. In Perseus, there is often an offset of $0.6-1$ kpc between the H$\alpha$ filaments and the young stars \citep{Canning2010, Canning2014}. The cold gas is also often found behind or along the peripheries of radio bubbles, suggesting a link between cooling and AGN activities \citep{Fabian2003, Tremblay2016, Russell2016}. 

Recent high-resolution numerical simulations have shown that momentum-driven AGN feedback can suppress radiative cooling and stimulate thermal instabilities \citep{Gaspari2012, PII, Meece2016, Yang2016a}. The velocities of the cold gas in simulations typically fall in the same range as observations $\sim 100-400$ km/s \citep{Prasad2015}, which is much lower than expected if the gas falls ballistically to the cluster center from a few to a few tens of kpc. 

The effect of ram pressure on cold clouds has been studied extensively in our own galactic halo \citep[e.g.][]{Heitsch2009}. In galaxy clusters, ram pressure is mostly studied in the context of ram pressure stripping of gas from galaxies moving through the ICM \citep{Tonnesen2008, Ruszkowski2014}. In this paper, we examine the effect of ICM ram pressure on the cold clouds in the centers of cool-core clusters. We argue that if the cold gas is slightly ``fluffier'' (with a lower average density) than previously assumed, when taking AGN wind into consideration, the effect of ram pressure can slow down the motion of free-falling cold clouds such that their velocities are in agreement with observations and simulations. In addition, ram pressure can cause a separation between cold filaments and newly formed young stars which explains the typical offset observed in Perseus. Furthermore, the offset combined with the line-of-sight velocity measured from emission lines can help discern between inflows and outflows. 

The paper is structured as follows: Section~\ref{sec:sim} briefly describes the simulation discussed in the paper. Section~\ref{sec:terminal} shows the velocities of cold clumps in simulations and in observations (Section~\ref{sec:terminal_a}), and analyzes the effect of ram pressure on the clump velocity under different assumptions about clump density and ICM velocity (Section~\ref{sec:terminal_b}). In Section~\ref{sec:separation_a}, we show the offset between cold clouds and young stars in simulations. In Section~\ref{sec:separation_b}, we analyze how ram pressure causes this offset, and discuss how to use the offset to understand the true motion of the cold gas. We summarize this work in Section~\ref{sec:conclusion}.

\section{The Simulation}\label{sec:sim} 
In this section we briefly describe the simulation analyzed in this work, which is the standard run in \citet{Li2015}.

We use the adaptive mesh refinement (AMR) code Enzo \citep{Enzo} with the Zeus hydrodynamic solver \citep{Zeus}. The smallest cell size is $\sim 244$ pc. Our earlier work without star formation \citep{PIII,PII} has a higher resolution, but the general results are converged including the physical properties of the cold clouds (except their sizes). We set up an isolated Perseus-like cool-core galaxy cluster initially in hydrostatic equilibrium. The momentum-driven AGN feedback is modeled with a pair of collimated and mass-loaded non-relativistic jets, powered by the accretion of cold gas surrounding the SMBH in the center of the cluster. Radiative cooling is computed based on a cooling table with a temperature floor of 300 K and a half-solar metallicity \citep{CoolingFunction, Metallicity}. Other important physical processes include self-gravity of the gas, star formation and stellar feedback \citep{CenOstriker}. 

The simulated cool-core cluster experiences cycles of gas condensation/AGN outbursts on $1-2$ Gyr timescales. At the beginning of each cycle, in the absence of cold gas and AGN feedback, the ICM cools radiatively and the cluster relaxes towards a classical cooling flow profile. The onset of a global cooling catastrophe in the cluster center turns on AGN feedback, which dredges up low entropy ICM to larger radii, triggering more condensation along the jet path. Some of the condensed gas turns into stars, and some falls to the cluster center and feeds the SMBH. Shock dissipation facilitated with mixing and adiabatic processes globally heats up the ICM of the cluster core, slowing down further condensation. Eventually, when star formation and stellar feedback consume all the cold gas, AGN feedback is shut off, allowing the ICM to cool again: the system enters the next cycle (see \citet{Li2015} for details). 

The simulation produces a wide range of features in general agreement with the observations, including the spatially extended filamentary multi-phase gas and star forming structures \citep{Tremblay2015, Donahue2015}, and the low velocities and velocity dispersions of the hot ICM \citep{Li2016, Hitomi2016}. In this work, we focus on the cold clouds \footnote{The cold gas in both simulations and real clusters exists in filamentary/clumpy structures (they appear to be more clumpy than filamentary in simulations likely due to the lack of magnetic fields). We refer to the cold structures as (cold) filaments, clumps and clouds almost interchangeably in this work.}, and in particular, how ram pressure affects their velocities and what are the implications and applications. 

\section{The Velocities of In-falling Cold Clouds}
\label{sec:terminal}

\subsection{The Velocities of Cold Clouds in Simulations and Observations}
\label{sec:terminal_a}
In this section, we present the velocities of cold clouds measured in our simulations \citep{Li2015} and compare them with the observations. 

The molecular gas in the centers of galaxy clusters has line-of-sight velocities and velocity dispersion of a few hundred km/s (typically $< 200-300$ km/s) measured from emission lines \citep{McDonald12, Tremblay2016, Vantyghem2016, Russell2016}. The observed velocities are much lower than the free-fall velocity, i.e., if the molecular gas falls ballistically from a few tens kpc, which we discuss in more detail in Section~\ref{sec:terminal_b}. The velocities of the molecular gas in simulations of \citet{Li2015} are consistent with the observations, and many of the cold clumps do form at radii of a few tens of kpc \citep{PII}. 

\begin{figure*}
\begin{center}
\includegraphics[scale=.28]{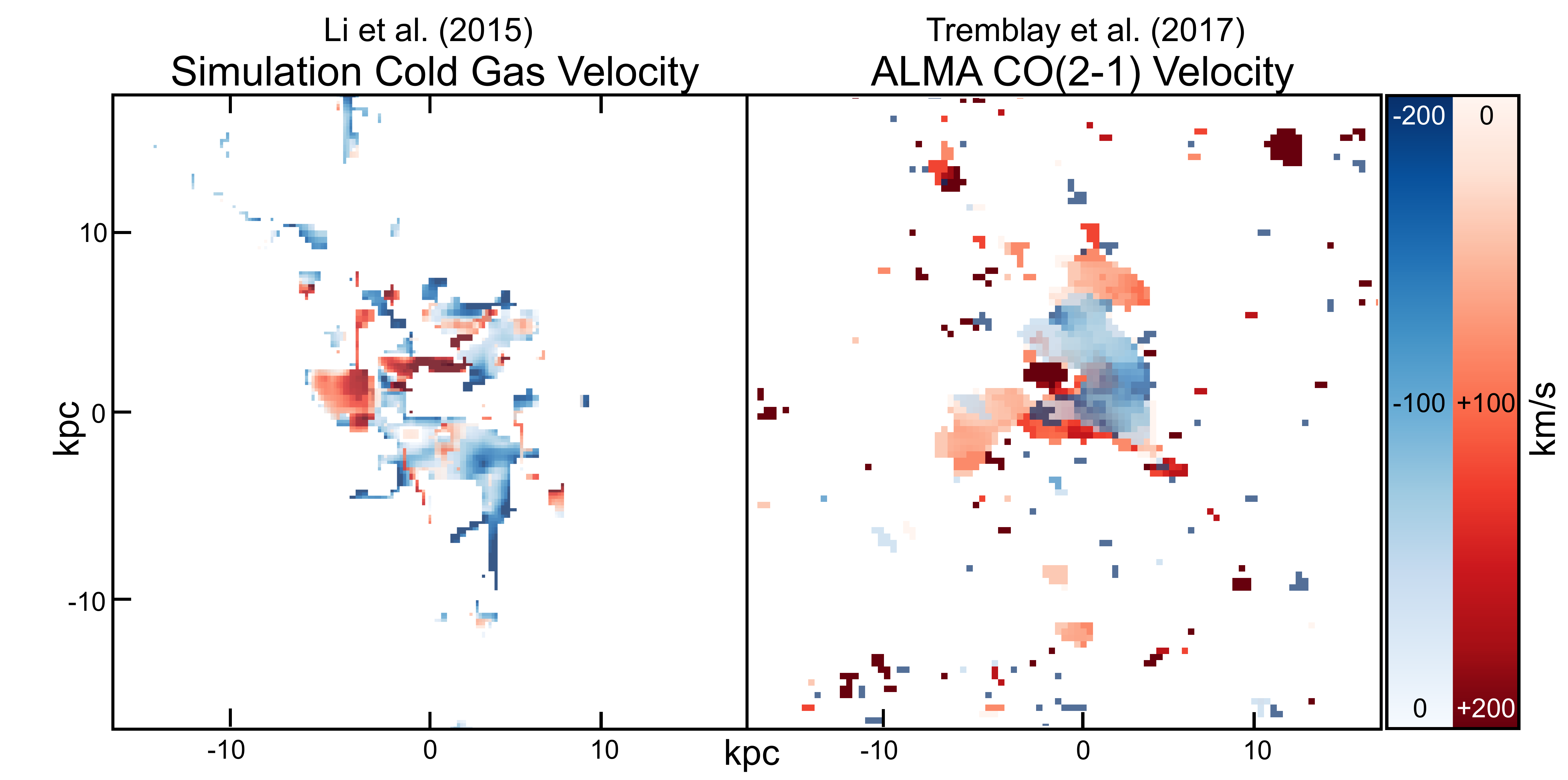}
\caption{Left: the line-of-sight velocity of the gas with temperatures below $10^4 K$ at t=0.72 Gyr in the standard simulation of \citet{Li2015}. The redshifted component with receding line-of-sight velocities is shown in red, and the blue-shifted component with positive line-of-sight velocities is shown in blue. Right: the line-of-sight velocity of the molecular gas in Abell 2597 based on the ALMA CO (2-1) observations (Tremblay et al. 2017 in preparation).
\label{fig:sim_obs_cloud}}
\end{center}
\end{figure*}

Figure ~\ref{fig:sim_obs_cloud} shows the line-of-sight velocities of the cold gas at t=0.72 Gyr in our simulation (left panel). We choose a snapshot with SFR $\sim 4.5 \rm M_\odot yr ^{-1}$ to compare with a nearby cool-core cluster Abell 2597 which has an IR SFR of $\sim 4-5 \rm M_\odot yr ^{-1}$ \citep{Donahue2007}. The right panel of Figure~\ref{fig:sim_obs_cloud} shows the velocities of molecular gas in the center of Abell 2597 based on the ALMA observations of CO(2-1) emission (Tremblay et al. in preparation). The simulations do not resolve the atomic and molecular physics. We select gas with temperatures below $10^4$ K to compare with the observed molecular gas. The cold gas is then decomposed into a redshifted (shown in red) component with receding line-of-sight velocities, and a blue-shifted component with positive line-of-sight velocities shown in blue. The observed molecular gas in Abell 2597 is decomposed and plotted in a similar fashion. In both simulations and observations, the cold gas shows complex velocity structures, with the blue-and red-shifted components sometimes over-laid or next to each other in projection. The typical velocities in both simulations and observations are below 200 km/s. 

\begin{figure}
\begin{center}
\includegraphics[scale=.48]{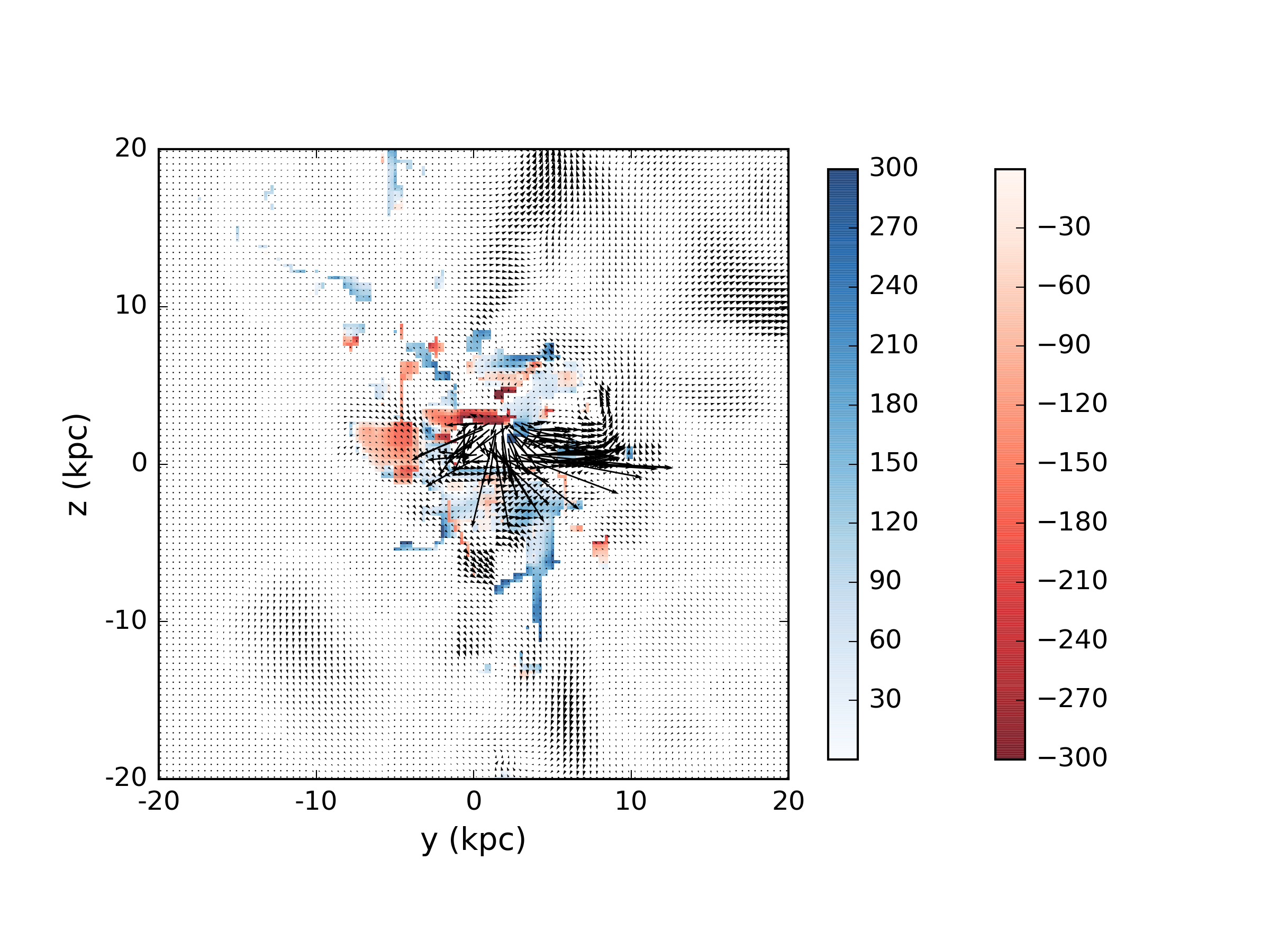}
\caption{The line-of-sight velocity of the gas with temperatures below $10^4 K$ at t=0.72 Gyr in the standard simulation of \citet{Li2015} (same as the left panel of Figure~\ref{fig:sim_obs_cloud}. Black arrows annotate the transverse velocity field of all the gas including both cold gas and hot ICM. Note that this figure uses the actual velocity range ($-300-300$ $km/s$) without saturation, which is wider than the range used in Figure~\ref{fig:sim_obs_cloud}.
\label{fig:sim_cloud}}
\end{center}
\end{figure}

In Figure~\ref{fig:sim_cloud}, the transverse velocity field is over-plotted as black arrows, which shows the velocities of all the gas within a slice through the cluster center projected on the y-z plane. The velocities of the hot ICM and the cold gas are not distinctively different, and are both rather low in most areas. In the very center of the cluster where the jets are launched, the velocities of the hot ICM can be rather high ($> 1000$ km/s, see Section ~\ref{sec:separation_b} for more discussion). When the high velocity outflow hits the cold gas, instead of uplifting the cold gas, it gets blocked and redirected, finding its path of least resistance. Even though the jets are launched along the z-axis (vertical direction) from their base, the actual outflow at this moment is channeled out mostly along the positive y-axis. 

\begin{figure}
\begin{center}
\includegraphics[scale=.38]{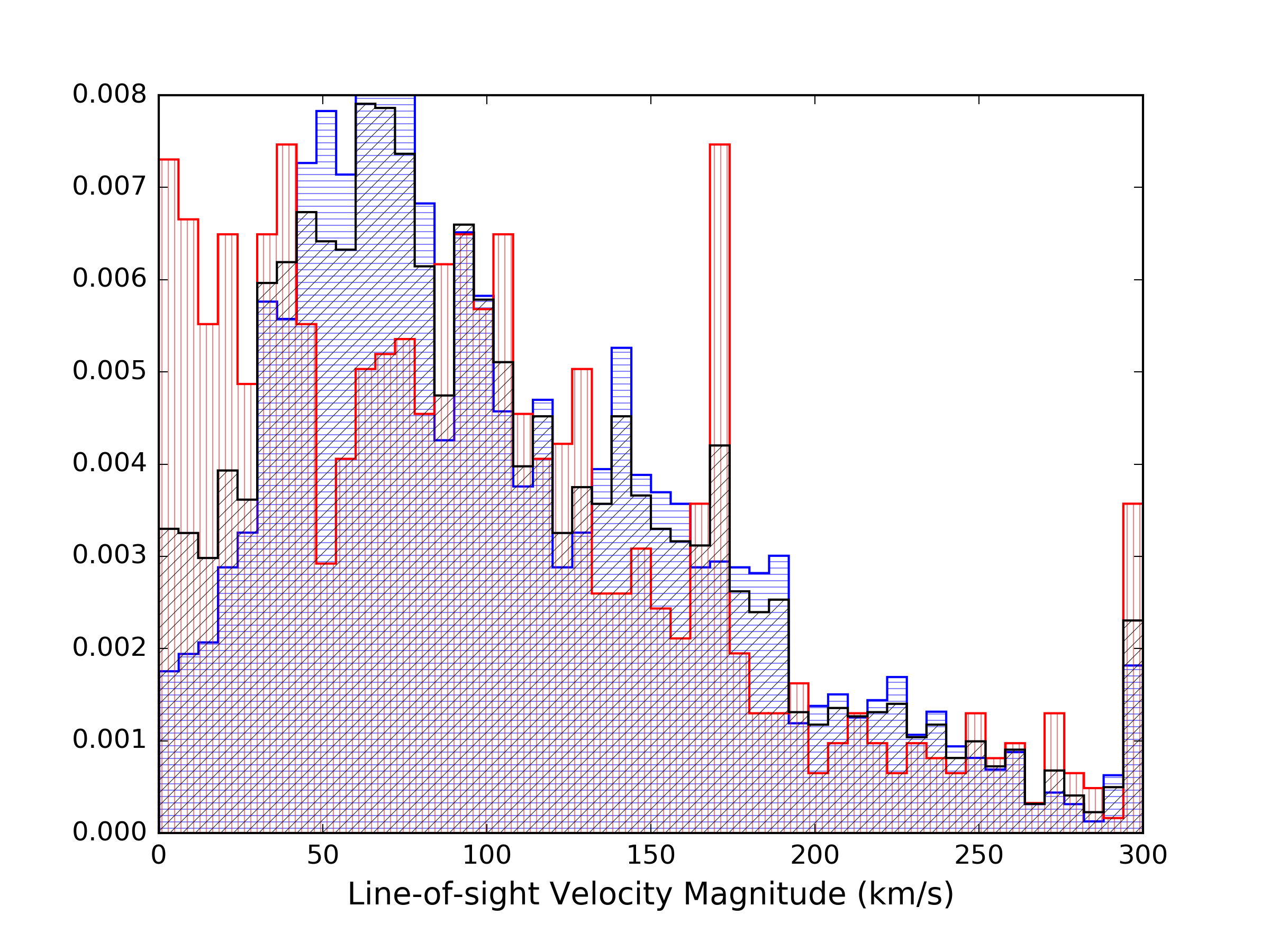}
\includegraphics[scale=.38]{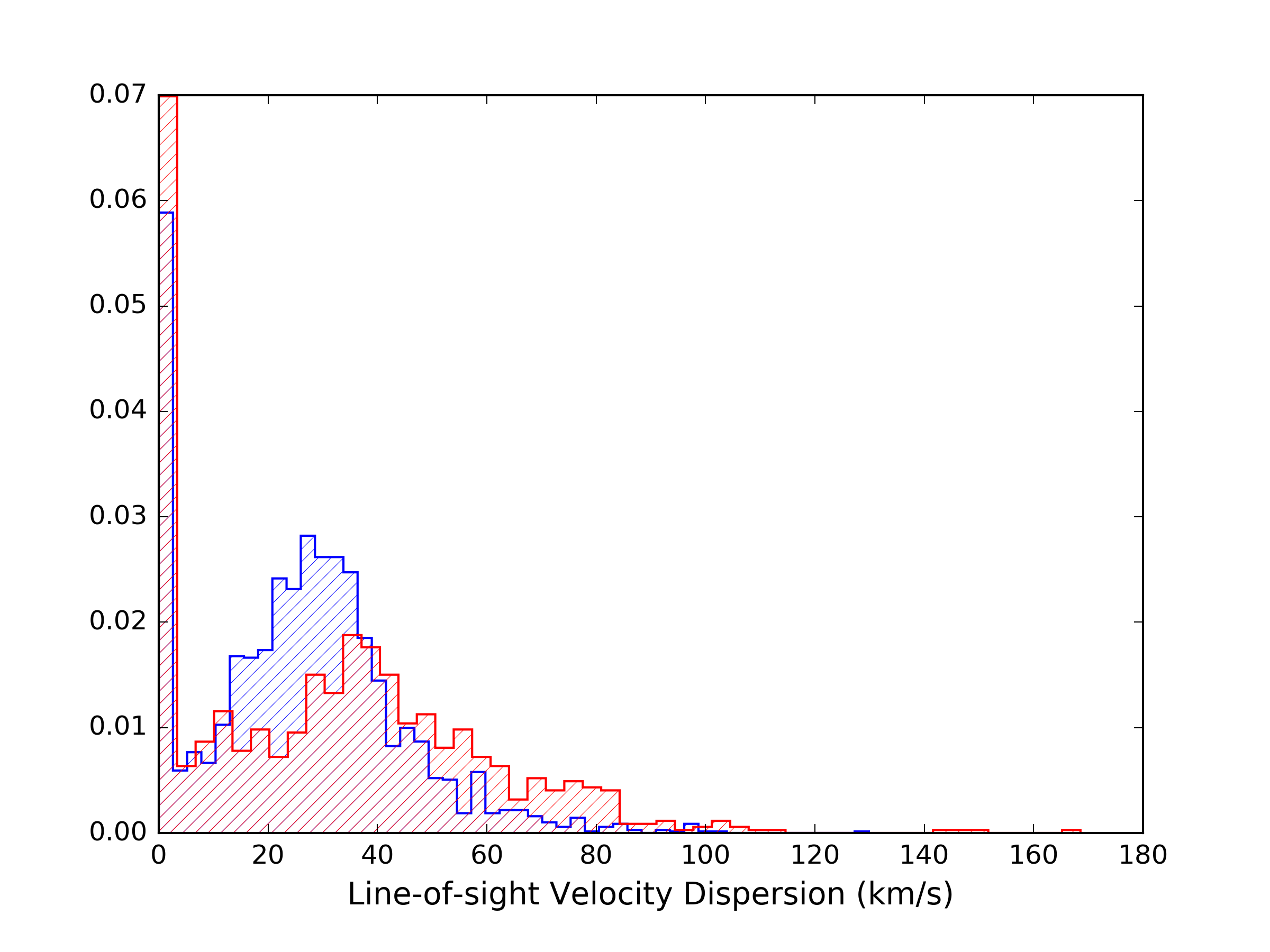}
\caption{Top: The line-of-sight velocity distribution of the gas with temperatures below $10^4 K$ at t=0.72 Gyr in the simulation, viewed along the x-axis. Red and blue show the red-shifted and blue-shifted gas, respectively. The distribution of the two components combined is shown in black. Bottom: the distribution of velocity dispersions along lines of sight of the same gas as the top panel. 
\label{fig:v_sigma}}
\end{center}
\end{figure}

The distribution of cloud velocities along lines of sight is shown in the top panel of Figure~\ref{fig:v_sigma}. The velocities of both components are below $360$ km/s. Most of the cold gas has velocities below 200 km/s. The line-of-sight velocity dispersion is also rather low, with a typical value of a few tens of km/s, the distribution of which is shown in the bottom panel of Figure~\ref{fig:v_sigma}.

In addition to the snapshot shown here, we have also looked at the velocity distribution at other times and found that the velocity range ($\sim 100-400$ km/s) is similar to what is found in \citet{Prasad2015} and is in general agreement with the observed range. The actual distribution of the velocities can be dependent on the simulation resolution and the model parameters (for example, the mass of the cluster, the width and the initial velocity of the jets). The focus of this work is not to reproduce the observed velocity distribution in simulations, but rather to use the simulations to gain insights as to why the velocities measured in both real and simulated clusters are different from simple theoretical calculations as we discuss in the following section.

\subsection{The Non-ballistic Motion of Cold Clouds}
\label{sec:terminal_b}

In this section, we compute the velocity of a cold cloud in-falling through the ICM towards the cluster center. As is shown in \citet{PII}, in simulations with momentum-driven AGN feedback, cold clouds often have positive radial velocities (moving outward) when they first form as the AGN jet triggers condensation in the marginally stable ICM. Precipitation can also happen when the ICM is thermally unstable \citep[see discussion in][]{Mark2016, Choudhury2016}. Alternatively, the cold gas may have formed at the bottom of the potential and been dredged up by radio bubbles \citep{Werner2010, Tremblay2012}. The different formation scenarios may lead to differences in the velocities of the outflowing molecular clouds. However, the focus of this work is not to decide which formation mechanism is correct or dominant. In all cases, cold clouds experience similar in-falling processes. That is, when the cold clouds form out of linear thermal instabilities of the ICM, or when they hit the turn-over point of their motion after decoupling from the AGN outflows or rising radio bubbles, they fall towards the center of the cluster due to gravity and zero initial radial velocity. The focus of this section is this in-falling motion of the cold gas.

We consider a cloud of mass $m$, length $l$ and a cross section of $\sigma$ moving radially and ignore the transverse motion. We assume that the cloud starts from a distance of $r$ with zero initial velocity. For simplicity, we first assume that the ICM is at rest and the gravitational acceleration $g$ does not change with radius. The acceleration of the cloud can be expressed as:

\begin{equation}
m\frac{dv}{dt}=m g - \rho_\mathrm{ICM} v^2 \sigma\, , 
\end{equation}
where $m=\rho_f \sigma l$, with $\rho_f$ being the mean density of the cloud. Thus we have:
\begin{equation}
\frac{dv}{dt}=\,g-\frac{f}{l}v^2\,,
\end{equation}
where $f=\rho_\mathrm{ICM}/\rho_f$ is the density contrast between the ICM and the cloud.

We solve for the velocity as a function of time:
\begin{equation}
v=\frac{2\sqrt{lg/f}}{1.0+e^{-2t\sqrt{fg/l}}}-\sqrt{lg/f}\,.
\end{equation}

Taking a typical ICM electron density of 0.04 $\rm cm^{-3}$ in the center of cool-core clusters \citep{Churazov2003, McDonald10}, and an average filament density of 2 $\rm cm^{-3}$ \citep{Fabian2008}, the terminal velocity of a cold cloud of length $l=2$ kpc is:
\begin{equation}
v_{terminal}=\sqrt{lg/f}\approx 962\, km\,s^{-1}.
\end{equation}

This velocity far exceeds that of any observed cold filaments. It is worth noting though that the length scale over which the terminal velocity is reached is actually larger than the largest radius where cold clouds are observed. The velocity only gets close to 800km/s after it has traveled a distance of 50 kpc, but in both simulations and observations, clouds that form at distances larger than 50 kpc are extremely rare \citep{McDonald10, Gaspari2012, PIII, Prasad2015}. Most cold clouds do not have a chance to actually reach their terminal velocity. However, the calculation above assumes a constant gravitational acceleration, which is a good approximation at a few tens of kpc. As the cold gas gets closer to the cluster center, the acceleration increases further due to the stellar potential of the BCG and the SMBH in the very center. As a result, the actual velocity of cold clouds could approach $\sim 900$ km/s in a realistic cluster potential even though the travel distance is short.

\begin{figure}
\begin{center}
\includegraphics[scale=.38]{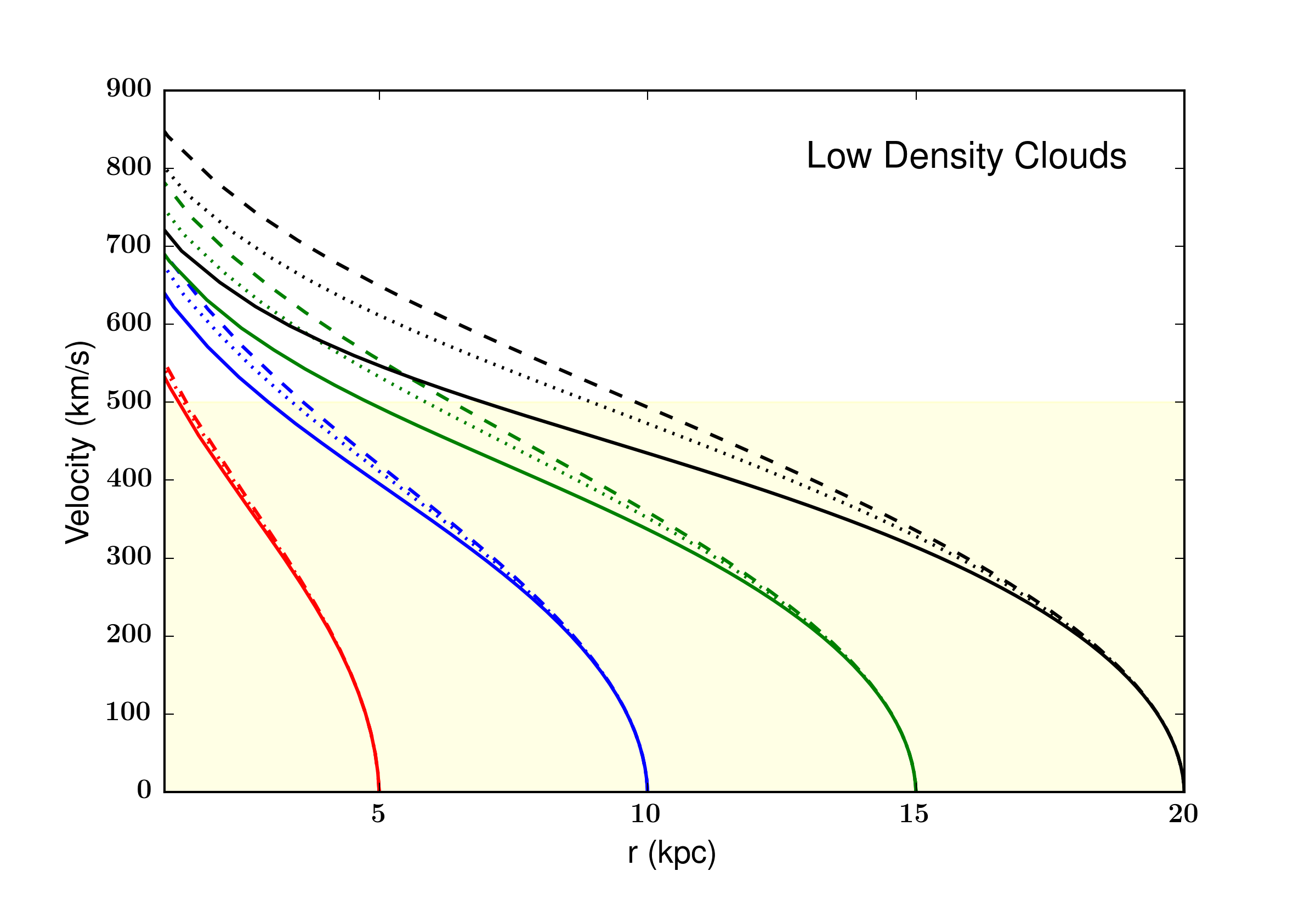}
\caption{The velocity of a cold cloud as a function of radius r if the cloud falls from $r = 5$, 10, 15 and 20 kpc. The cloud is assumed to have a length $l=2$ kpc and zero initial velocity. The dashed lines show the ballistic trajectory. The dotted lines take ram pressure into consideration. The solid lines assume the cold clouds are ``fluffy'', with an average density $1/3$ of the previous assumption. The yellow shaded area denotes the typical range of filament velocities observed in cool-core clusters.
\label{fig:v_fluffy}}
\end{center}
\end{figure}

Figure~\ref{fig:v_fluffy} shows the velocity of a cold cloud that falls to the center of a cluster with a Perseus-like gravitational potential, where an NFW dark halo dominates at $r>10$ kpc and the BCG and SMBH dominate the center \citep{Mathews, P1}. The cloud is released with no initial velocity from $r=$5, 10, 15 and 20 kpc. The dashed lines are the ballistic trajectories and the dotted lines take into consideration of the effect of ram pressure. The highlighted yellow area denotes the typical observed velocities of cold gas in the center of cool-core clusters \citep{McDonald12}. Ram pressure only slows down the clouds slightly and the final velocities of the clouds are still too high regardless of their initial location. 

Obviously not all clouds are moving at their highest velocities. Any observation is only a snapshot when many of the clouds are likely still accelerating. This partially explains why the observed velocity of the cold clouds is typically lower than the terminal velocity calculated above. In addition, when the clouds are not oriented perfectly radially, the cross section increases and the velocity of the clouds can be smaller. However, we should expect to see some cold clouds moving at velocities on the order of $1000$ km/s if they simply fall to the SMBH ballistically from a few to a few tens of kpc. The ``high velocity clouds'' are missing in both the observations and simulations (Figure~\ref{fig:sim_cloud} and \ref{fig:v_sigma}), which motivates us to re-examine the assumptions made in the calculations above. 

First, it is possible that we underestimate the effect of ram pressure because the average density of the filament is actually lower than the value we use. \citet{Fabian2008} estimated the average filament density to be $2 \rm cm^{-3}$ by dividing the total H$_2$ mass (from CO observations) by the volume of the filaments (estimated from optical images). The total volume of the actual filament as a coherent moving structure may be larger. Cold clumps in simulations are surrounded by layers of gas with intermediate temperatures \citep{PII} (though the thickness of the layers may be resolution dependent). The observed optical filaments also co-reside with soft X-ray features \citep{Fabian2006}. If we assume that typical cold filaments are surrounded by a warm-hot coat that reduces the average filament density to 1/3 of the value previously used, the effect of ram pressure becomes more significant (solid lines in Figure~\ref{fig:v_fluffy}). However, the velocities near the center are still too high. 

\begin{figure}
\begin{center}
\includegraphics[scale=.38]{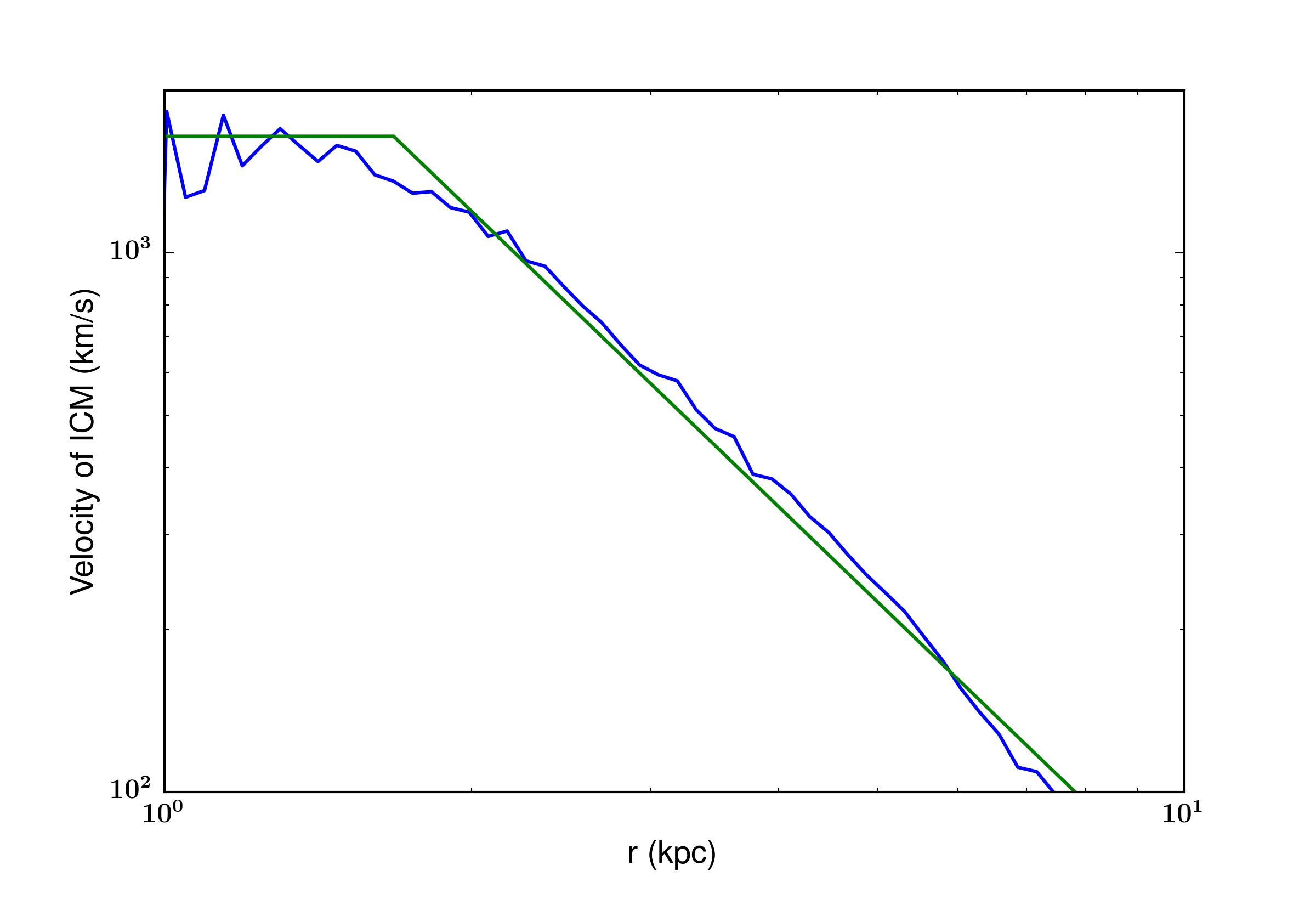}
\caption{The velocity profile of AGN-driven wind. We select the ICM with temperatures above $10^7$ K as a simplistic way of selecting the AGN outflow. The blue line shows the velocity profile averaged over the entire simulation, and the green line is a simple fit to the curve, with $v_{wind}=1.6\times 10^3$ km/s at $r< 1.7$ kpc and $v_{wind}\propto r^{-1.82}$ at $r>1.7$ kpc. The yellow shaded area denotes the typical range of filament velocities observed in cool-core clusters.
\label{fig:wind}}
\end{center}
\end{figure}

Second, one factor that is not considered in previous calculations is the motion of the ICM. Because AGN in cool-core clusters is almost always on in the simulation and in observations \citep{Birzan2004, Dunn2006}, the cold clumps are not moving in ICM that is at rest, but rather, moving against AGN-driven wind. We measure the volume-weighted velocity of gas at temperatures above $10^7$ K (as a simplistic way of selecting the AGN outflow) and take the time-averaged velocity as the average wind velocity, shown in Figure~\ref{fig:wind}. The wind is only strong in the very center of the cluster, and wind velocity falls below 100 km/s at $r=8$ kpc, consistent with what is shown in Figure~\ref{fig:sim_cloud}. This is because kinetic energy is dissipated via strong shocks very quickly \citep{Li2016}. We use a simple formula to fit the wind velocity profile: $v_{wind}=1.6\times 10^3$ km/s at $r< 1.7$ kpc; $v_{wind}\propto r^{-1.82}$ at $r>1.7$ kpc. Note that this profile is azimuthally averaged wind velocity averaged over the duration of the simulation of $\sim 6.5$ Gyr. The exact shape of the profile is likely dependent on the model parameters. In real clusters such as Perseus, the bubbles appear more isotropic than the bi-polar jets in the simulations, likely due to cluster weather \citep{Heinz2006} and/or re-orientation of the jets \citep{Babul2013}. The focus of this work is not the exact azimuthal and radial profiles of the wind. Instead, we aim to demonstrate that the wind has a nontrivial effect on the velocity of the cold gas close to the center of the cluster.

\begin{figure}
\begin{center}
\includegraphics[scale=.38]{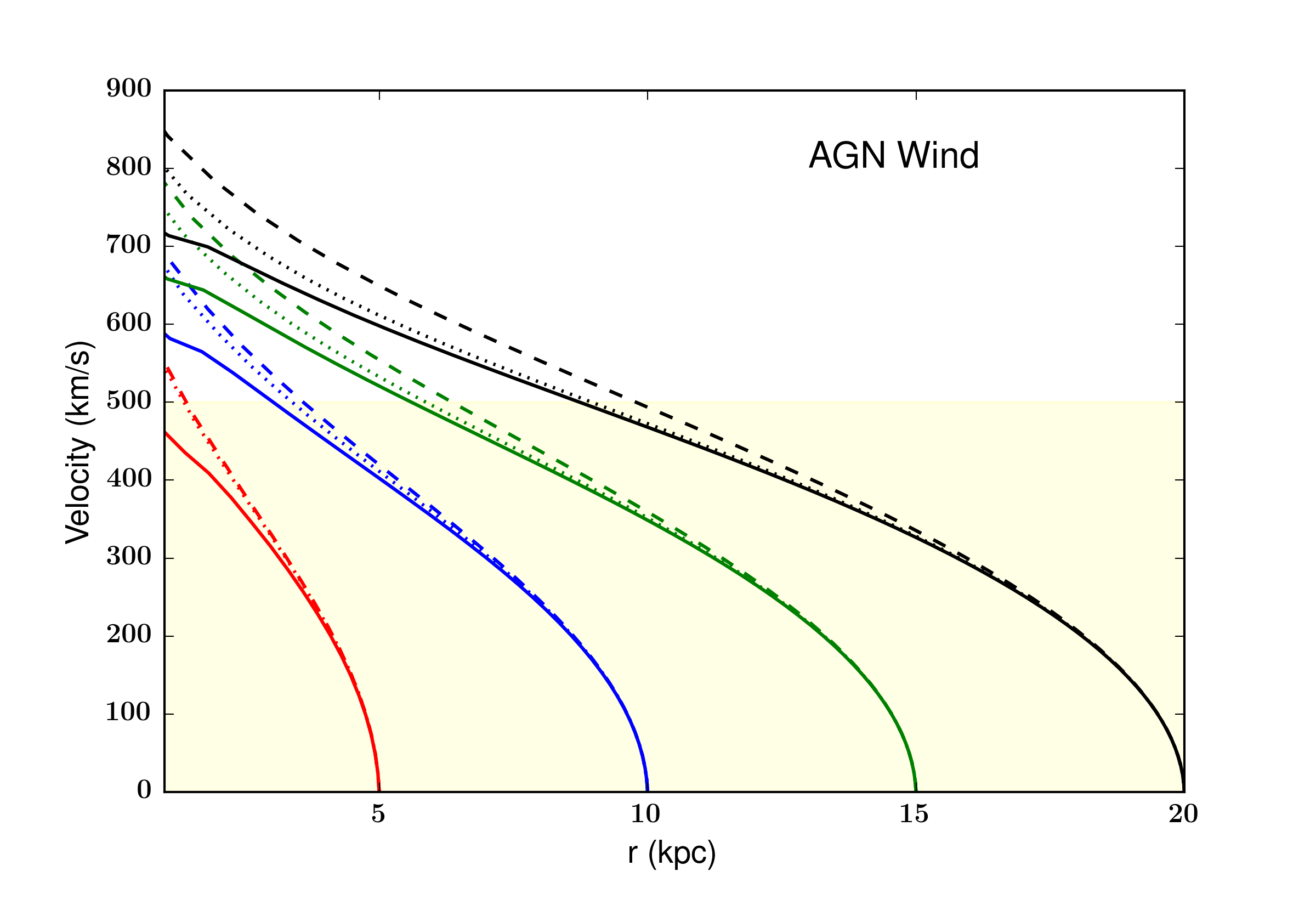}
\caption{The velocity of a cold cloud as a function of radius r if the cloud falls from $r = 5$, 10, 15 and 20 kpc, similar to Figure~\ref{fig:v_fluffy}. The dashed lines show the ballistic trajectory. The dotted lines take ram pressure into consideration assuming the ICM is at rest. The solid lines take into consideration the effect of AGN winds, with the wind profile shown in Figure~\ref{fig:wind}. The yellow shaded area denotes the typical range of filament velocities observed in cool-core clusters.
\label{fig:v_wind}}
\end{center}
\end{figure}

When we add this wind velocity correction to the original calculation, still assuming a filament density of $2 cm^{-3}$, the resulting velocities are suppressed, especially close to the center where the wind velocity is the highest (solid lines in Figure~\ref{fig:v_wind}). However, the velocities are still too high compared with observations. This means that AGN wind alone does not provide enough force to slow down the cold gas if we use the commonly used filament density. 

\begin{figure}
\begin{center}
\includegraphics[scale=.38]{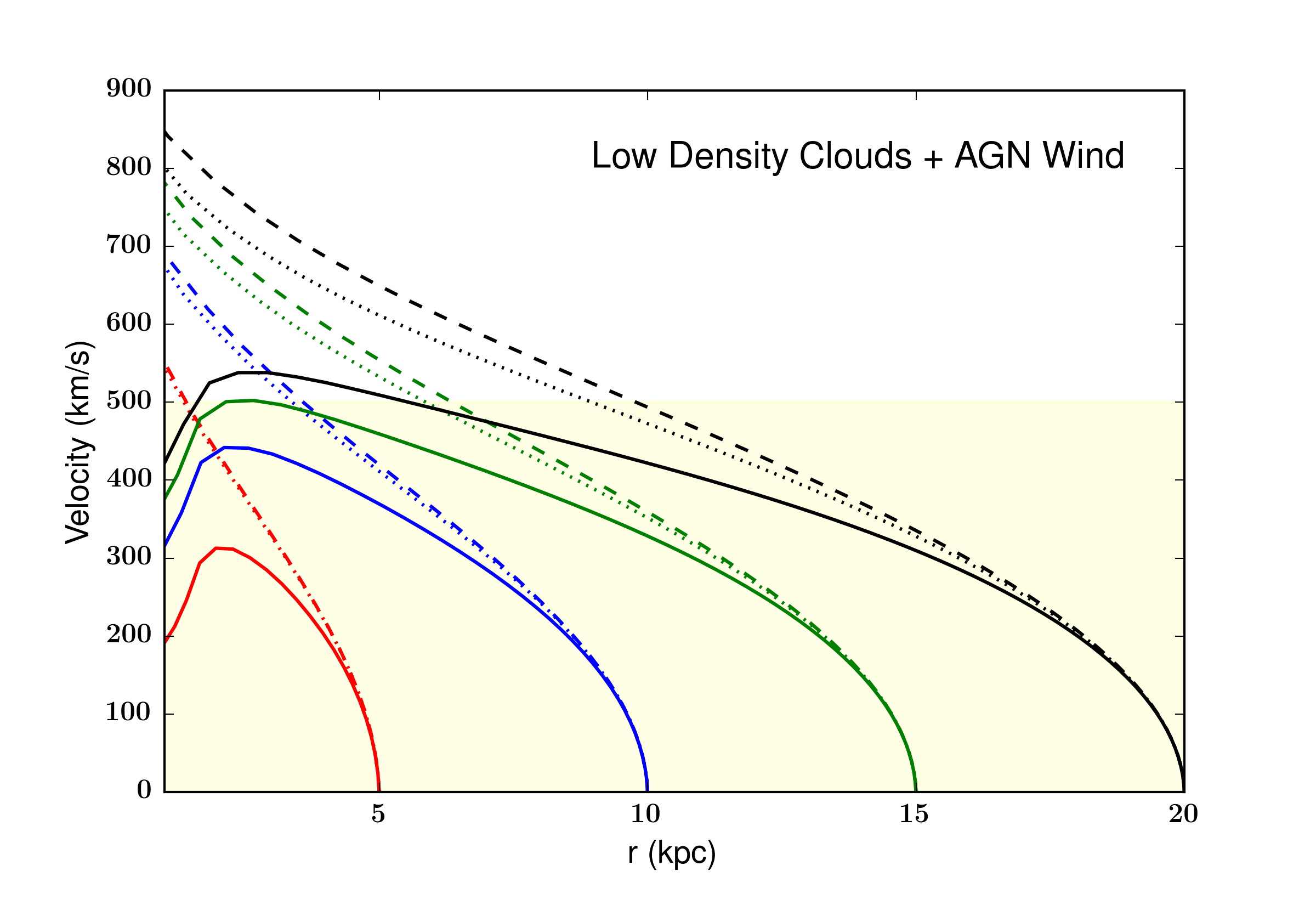}
\caption{The velocity of a cold cloud as a function of radius r if the cloud falls from $r = 5$, 10, 15 and 20 kpc, similar to Figure~\ref{fig:v_fluffy} and ~\ref{fig:v_wind}. The dashed lines show the ballistic trajectory. The dotted lines take ram pressure into consideration assuming the ICM is at rest. The solid lines take into consideration the effect of AGN winds and also assume ``fluffy'' clouds with $1/3$ of the commonly assumed average density. The yellow shaded area denotes the typical range of filament velocities observed in cool-core clusters.
\label{fig:v_fluffywind}}
\end{center}
\end{figure}

When we consider the AGN wind, and assume that the average density of the moving filaments is lower because of the warm-hot layer surrounding the cold gas, then the velocities of the filaments are in agreement with the observations (solid lines in Figure~\ref{fig:v_fluffywind}). 

Besides reducing the velocity, ram pressure can also strip or even shred a cold cloud. The clouds moving at higher velocities are more prone to ram pressure stripping. The preferential destruction of high velocity clouds also helps explain the lack of cold gas moving at very high velocities in both observations and simulations.

In addition to ram pressure, a cold cloud can be further slowed down by exchanging material with the ICM perpendicular to its direction of motion. As \citet{PII} shows, when cooling instability happens, as the temperature of the newly formed clump decreases, $t_{cool}$ exceeds sound crossing time at some point, and cooling is no longer isobaric. As a result, most of the cold clouds have internal pressure lower than the ICM pressure, thus drawing more hot gas to cool onto the cloud in a ``mini cooling flow''. Ram pressure stripping and stellar feedback can remove some of the cold gas from the cloud, which gets mixed to the hot ICM. The exchange of material results in an exchange of momentum between the moving cloud and the ICM at rest, equivalent to increasing the ram pressure cross section. Future simulation work with better resolution is needed to quantify the effect of these processes.

Magnetic fields in the ICM, which are not included in the simulation, can further enhance the drag force on the clouds (though the effect is weak for high $\beta$ plasma) \citep{McCourt2015}.

\section{The Offset between Filaments and Young Stars}
\label{sec:separation}

\subsection{Star-Filament Separation in Simulations and Observations}
\label{sec:separation_a}

\begin{figure}
\begin{center}
\includegraphics[scale=.25,trim=0.3cm 0.2cm 0.2cm 0.2cm, clip=true]{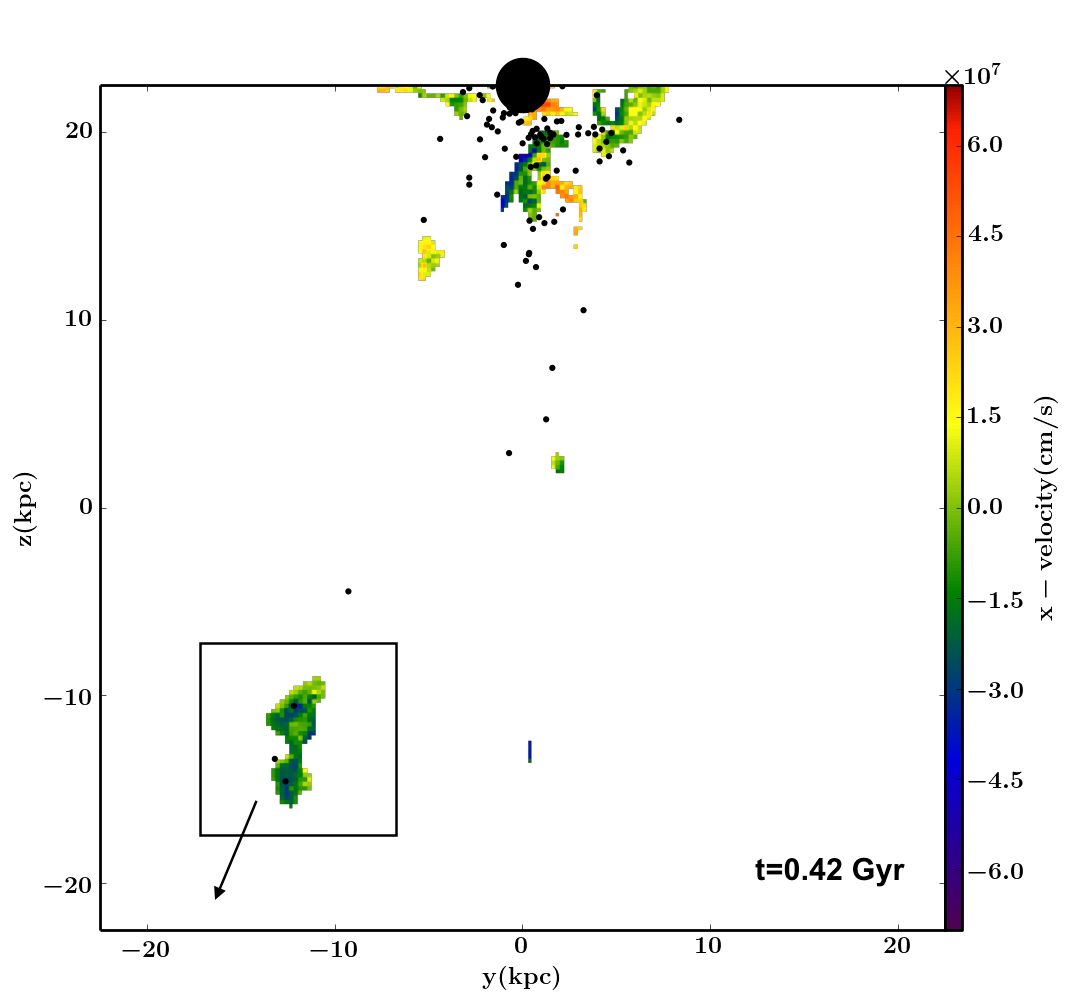}\\
\includegraphics[scale=.25,trim=0.3cm 0.2cm 0.2cm 0.2cm, clip=true]{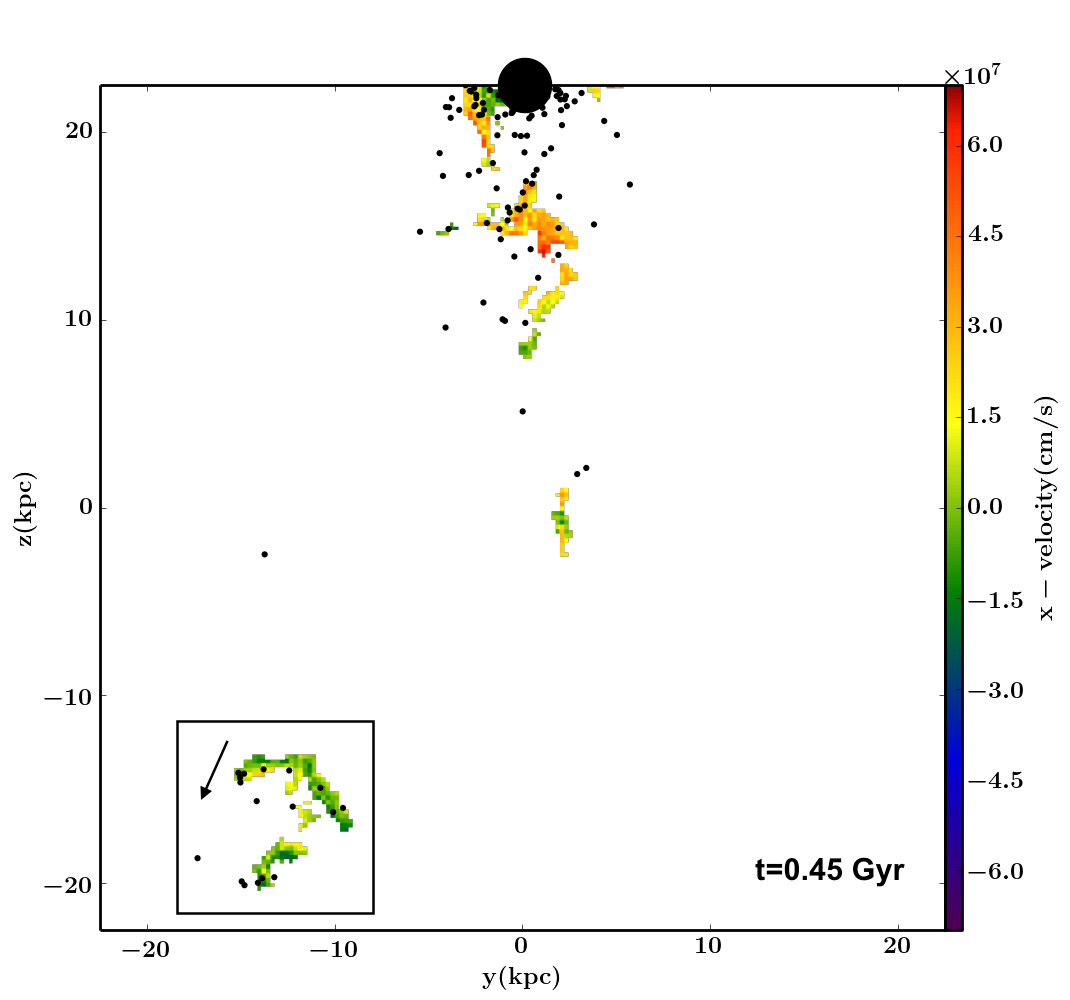}\\
\includegraphics[scale=.25,trim=0.3cm 0.2cm 0.2cm 0.2cm, clip=true]{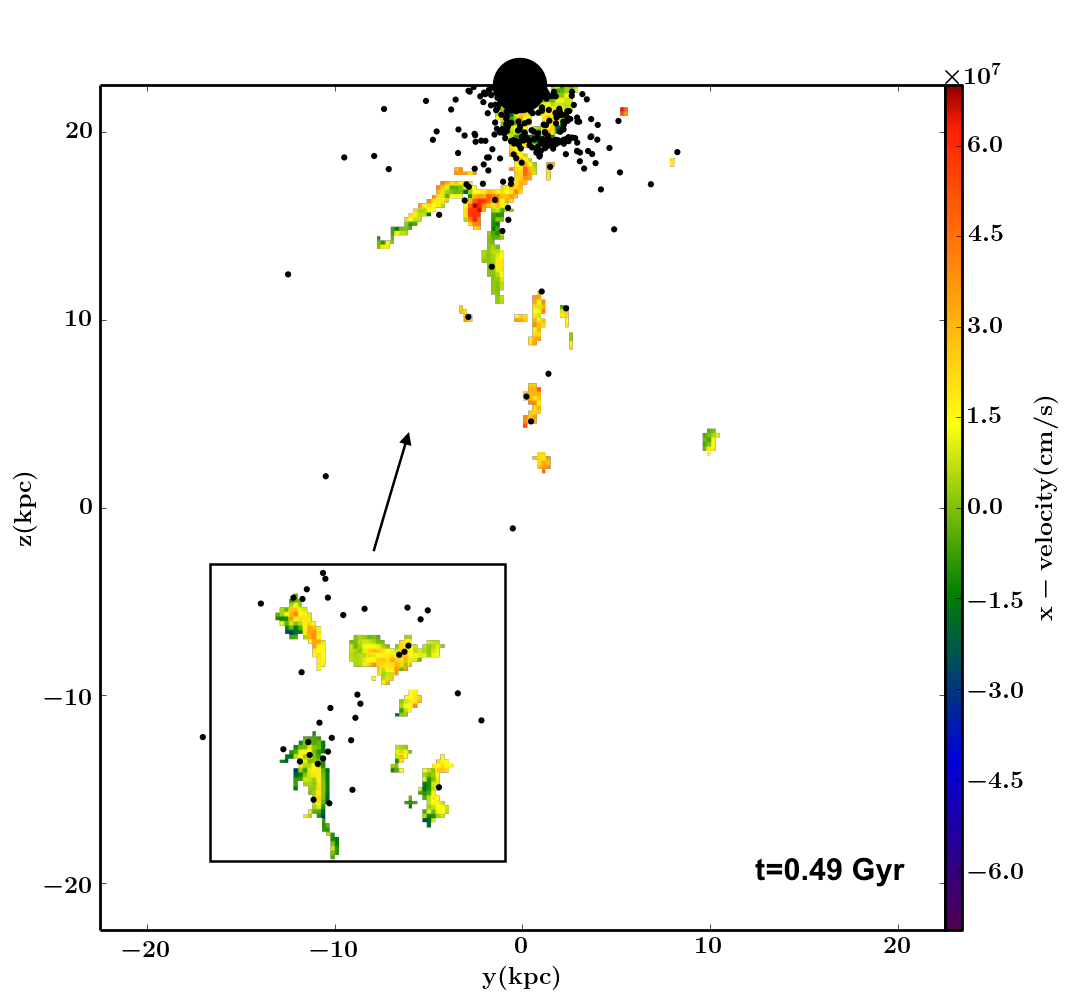}
\caption{Young stars (black dots) and cold gas in a rectangular region of $(50 kpc)^3$ near the center of the cluster at $t=0.42$, 0.45, and 0.49 Gyr in the standard simulation of \citet{Li2015}. Color corresponds to line-of-sight velocities weighted by the $H\alpha$ luminosity. The big black circle represents the SMBH in the center of the cluster, which is powering AGN jets in the vertical direction, pointing downwards in the image. The small rectangle marks the cold clouds and young stars whose motion is discussed in Section~\ref{sec:separation_a}. The arrow points to the direction of their transverse motion. 
\label{fig:offset_sim}}
\end{center}
\end{figure}

Observations of the young star clusters in the outer halo of NCG1275, the central galaxy in the Perseus cluster, show typical projected spatial offset from the H$\alpha$ filaments of $0.6-1$ kpc \citep{Canning2010, Canning2014}. The offset between young stars and cold molecular gas is also seen in the simulations in \citet{Li2015}. Figure~\ref{fig:offset_sim} shows such an example. A cold cloud forms as the low entropy ICM is uplifted by the AGN jets from smaller radii \citep{PII, Mark2016, McNamara2016}. At $t=0.42$ Gyr, stars start to form (top panel of Figure~\ref{fig:offset_sim}) inside the outward-moving cold cloud. At the time, the line-of-sight velocity (weighted by collisional $H\alpha$ emissivity) is blue-shifted. At $t=0.45$ Gyr, as the cloud approaches its apocenter, there appears to be a systematic separation between the cold gas and the young stars, with the stars leading the gas by $\sim$ 1 kpc (middle panel of Figure~\ref{fig:offset_sim}). At $t=0.49$ Gyr (bottom panel of Figure~\ref{fig:offset_sim}), the cold gas, which has been shredded into a few separate pieces (some of them grew bigger via the ``mini cooling flow'' discussed in Section~\ref{sec:terminal_b}, is falling back towards the SMBH due to gravity. The stars have moved to the opposite side of the cold gas and are leading again. The cold gas is mostly red-shifted now. The clouds and stars are in the sixth octant (between the observer and the y-z plane).

Note that the simulation is shown mainly to demonstrate the point, but is not intended to reproduce realistic star formation and stellar feedback on small scales. As discussed in more detail in \citet{PII}, the size of individual clouds is resolution limited, and their shape is more ``clumpy'' and not as ``filamentary'' as the clouds in Perseus, likely due to the lack of magnetic fields in the simulations \citep{Wagh2013}. The stars are also represented by massive star particles in the simulation with a minimum mass of $\rm 10^6M_\odot$, which is more massive than the observed young star clusters in Perseus. Nonetheless, we see offset between young stars and cold gas both in simulations and in observations, which we argue is caused by similar physical processes discussed in detail in Section~\ref{sec:separation_b}.

\subsection{Ram Pressure Causing Young Stars to Move Away from Filaments}
\label{sec:separation_b}
The observed filaments stretch out predominantly radially \citep{Lim2009, Canning2010} and their motion is predominantly radial as well \citep{Prasad2015}. Some filaments can be moving at an angle with respect to their direction of motion. In this section, we discuss how ram pressure on these cold filaments can cause a spatial separation between filaments and the newly formed stars, and how the separation can help us understand the motion of the filaments.

\begin{figure}
\begin{center}
\includegraphics[scale=.3]{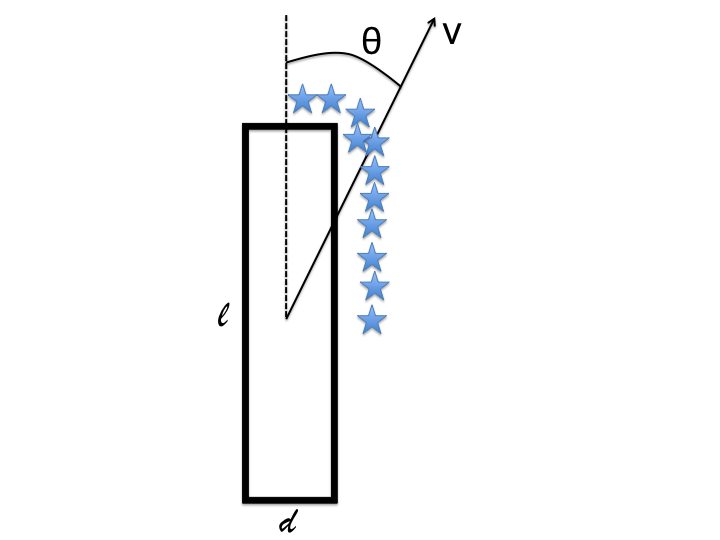}
\caption{An illustration of a cold filament forming stars as it moves through the ICM. We compute the separation between the filament and the young stars in Section~\ref{sec:separation_b}.
\label{fig:drawing}}
\end{center}
\end{figure}

For simplicity, we consider a cylindrical filament of length $l$ and diameter $d$. We assume the ICM is at rest, and the filament moves through the ICM at projected velocity $v$ at an angle $\theta$ measured from the major axis of the filament (see Figure~\ref{fig:drawing} for illustration). Assuming a group of stars form at $t=0$, the ram pressure exerts a force on the filament, but not the stars, causing the filament to move away from the stars:

\begin{equation}
\rho_\mathrm{ICM}v_{\perp}^2\,l \, d=\rho_f \, \pi\left(\frac{d}{2}\right)^2\, l \,\frac{d^2s}{dt^2}\,,
\end{equation}

\noindent where $s$ is the separation between the filament and the stars, $v_{\perp}=v\,cos\theta$ is the perpendicular velocity.

\begin{figure}
\begin{center}
\includegraphics[scale=.48]{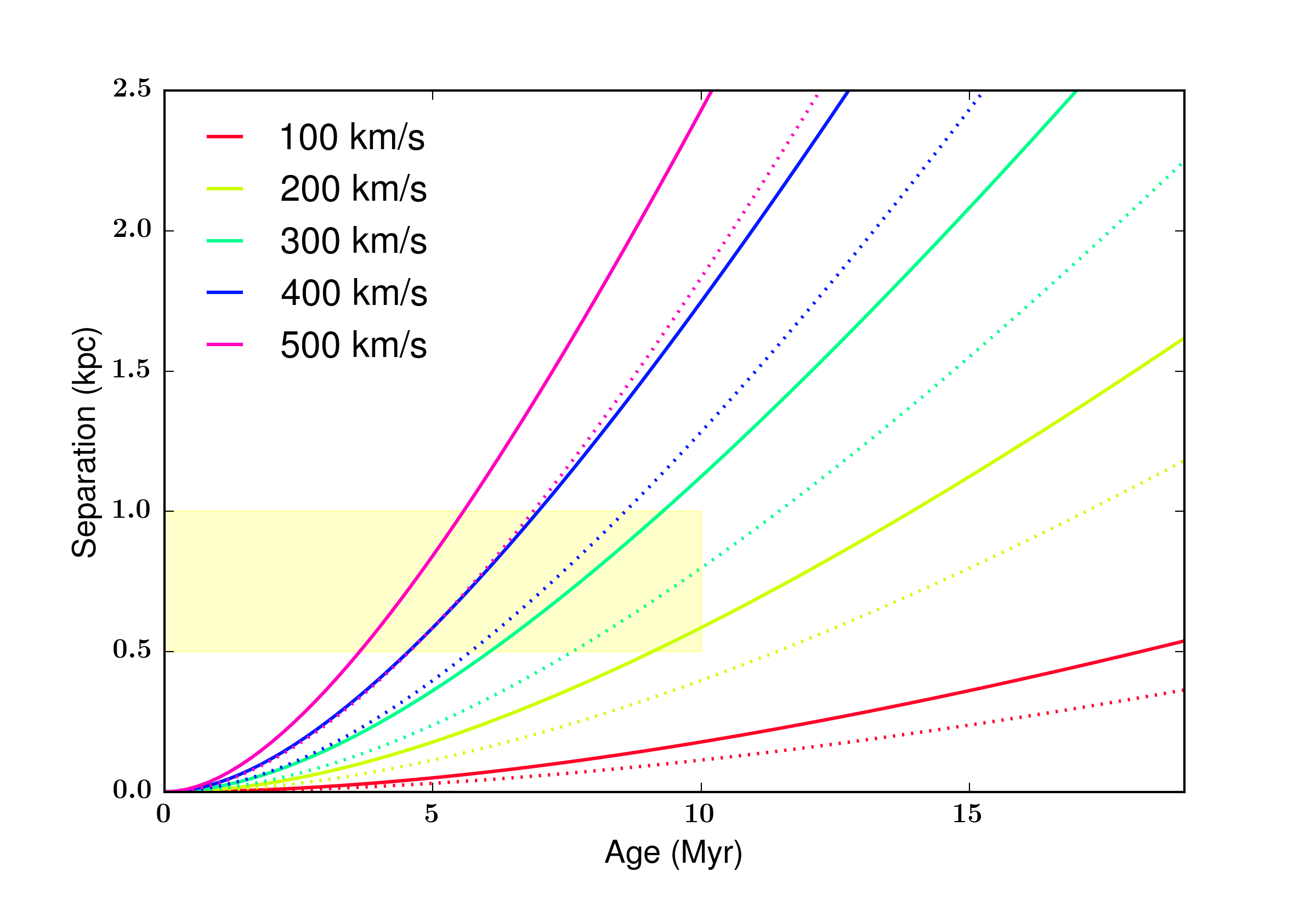}
\caption{The spatial separation between a cold filament and the newly formed young stars as a function of the stellar age for filaments moving at different initial velocities $v_{0}$ shown in different colors. The dashed lines use the canonical $2 \rm cm^{-3}$ for the average filament density, and the solid lines assume ``fluffy'' filaments with $1/3$ of the canonical density. The yellow shaded area denotes the optimum combination of separation and stellar age for easy detection (see Section~\ref{sec:separation} for discussion).
\label{fig:sep}}
\end{center}
\end{figure}

We can solve for the separation $s$ as a function of the initial velocity $v_{0}$, and time $t$ which is also the age of the stars:
\begin{equation}
s= v_{0}t-\frac{1}{A}\left(ln\left(At+\frac{1}{v_{0}}\right)+ln\left(\frac{1}{v_{0}}\right)\right)\,,
\end{equation}

\noindent where $A=\frac{4\rho_\mathrm{ICM}}{\pi d \rho_f}$. Figure~\ref{fig:sep} shows the star-filament separation $s$ as a function of the age of the stars for filaments moving at different initial perpendicular velocities $v_{0}$ (the stars will continue to move at velocity $v_{0}$). The dotted lines show the results when we assume a filament density of 2 $\rm cm^{-3}$ and a diameter of $d=100$ pc (\citet{Fabian2008} uses 35 pc as the characteristic radius of filaments in Perseus). The solid lines assume ``fluffy'' filaments with an average density of $2/3\, \rm cm^{-3}$ and a diameter of $\sqrt 3 d$. The yellow shaded area denotes the typical age of the young stars and the typical separation between young stars and filaments in Perseus. It is also the optimum combination of stellar age and separation for detection. O stars have a lifetime of up to about 10 Myr, so when star clusters are older than 10 Myr, they become much dimmer in the UV and thus harder to identify. $0.5-1$ kpc is a separation that is large enough to be easily visible, but not too large for it to become difficult to associate the stars with the filament. As is shown in Figure~\ref{fig:sep}, for a wide range of velocities ($\sim 200-500\, \rm km/s$), ram pressure can result in a separation between young stars and the filament that is easily detectable. The separation also becomes larger when assuming a lower average density for the filament as one would expect (solid lines compared with dotted lines). When we use a density of 2 $\rm cm^{-3}$, a velocity higher than $200\, \rm km/s$ is needed for the separation to be detectable. Note that this is only the perpendicular component of the velocity projected onto the sky ($v_{\perp}=v_f sin\phi\,cos\theta$ with $\phi$ being the angle between the velocity of the filament $v_f$ and the observer's line-of-sight), so the actual velocity would have to be even higher.

In addition to ram pressure, stellar feedback can also cause an apparent separation between young stars and filaments. If young stars form preferentially along one side of the filament, feedback from Type II SN may destroy the local molecular gas. The result of this process is a string of young star clusters next to the residual filament. It is possible that ram pressure itself can enhance star formation on the leading side of the filament as it moves through the ICM. Thus the leading side of the filament will be preferentially destroyed by Type II SN. In this case, stellar feedback will further enhance the separation between young stars and filaments. 

Observationally, it is often rather difficult to determine whether molecular gas is moving away from the cluster center in an outflow or falling back to feed the SMBH based on the line shifts only \citep{ALMA1, ALMA2, Vantyghem2016}. Only in rare cases can we claim with confidence that a cloud is falling onto the SMBH, e.g. when the cloud is seen in absorption \citep{Tremblay2016}. The star-filament separation can inform us of the true motion of the cold gas and thus help distinguish between inflows and outflows. Because stars are almost always leading (except very briefly at the apocenter), once we see an offset between the young stars and the filament, we know the general direction of the proper motion of the star-filament system. This information combined with the line-of-sight velocity measured from emission lines will tell us how the filament is moving through the ICM, in particular, whether it is moving out or falling back towards the cluster center. For example, if we observe a separation shown in Figure~\ref{fig:drawing}, and the cluster center is to the upper right of the filament, then the filament is falling towards the center of the cluster. If the lines are blue-shifted, then the filament is located further away from the observer than the center of the cluster; if the lines lines are red-shifted, the filament is falling towards the center from between the cluster center and the observer, which corresponds to the situation in the right panel of Figure~\ref{fig:offset_sim}. If the cluster center is to the lower left of the filament, then a redshifted filament would be moving away from the observer in an outflow, and a blue-shifted filament would be moving out towards the observer, which corresponds to a mirror image of the middle panel of Figure~\ref{fig:offset_sim}. 

The velocity of the cold gas is predicted to be predominantly radial in the main hypotheses that have been discussed in the literature; i.e., independent of whether the cold gas is condensing out of the ICM, is uplifted by the AGN bubbles, or forming on the interface of the jet and the ICM. This method of determining the velocity of the molecular gas is thus not restricted to the choice of any particular model. With enough sample of such measurements, we can try to distinguish between different AGN feedback models that produce extended filaments in different ways. Using the statistics of the measurement of offset, we can also put better constrains on the average density of the filaments.

\section{Conclusion}
\label{sec:conclusion}
We have discussed two phenomena related to the cold filaments in cool-core galaxy clusters that are possibly related to the effect of ram pressure from the ICM: the non ballistic motion of the cold gas, and the offset between cold filaments and young stars.

We measure the velocities of the cold gas in our numerical simulations and compare them with observations. We find that the velocities and velocity dispersions of the cold clouds in simulations fall in the same range as the observed ones (with a typical magnitude $< 200-300$ km/s), much lower than what is expected if the clouds fall to the cluster center ballistically. If we assume an average filament density of $2 cm^{-3}$ based on the apparent size of typical H$\alpha$ filaments in Perseus, the ram pressure of the ICM does not slow down the filaments enough. When we consider the effect of AGN wind blowing against the in-falling clouds in the center of the cluster ($r<10$ kpc) by applying an average wind profile measured from the simulation, we find that the cold filaments can be slowed down more. However, the desired velocities are achieved only when we also assume that the filaments are also ``fluffier'', i.e., if the cold filaments are moving with layers of warmer gas. This is supported by the simulations and the observed spatial correlation between H$\alpha$ filaments and soft X-ray features. The exchange of material between the filament and the ICM may further slow down its motion. 

An offset (typically of half to one kpc) between young stars and H$\alpha$ filaments is seen in the center of the Perseus Cluster. We observe similar offset between young stars and cold clouds in our simulation. We argue that this offset can be caused by ram pressure. The ICM only exerts ram pressure on the cold gas but not stars. As the whole structure moves through the ICM, the young stars that formed in the filament may appear to move away from the filament as the stars age. Because stars are always moving ahead of the cold cloud, the observed offset can inform us of their direction of motion projected onto the plane of the sky. This information, combined with the line-of-sight velocity obtained from emission line measurements, can give us the 3D velocity of the cold gas, and allow us to infer whether the gas is moving out as part of an outflow or falling back to the cluster center.

\acknowledgments

We thank Megan Donahue, Michael McDonald, Brian McNamara, Greg Bryan and Lee Hartmann for useful discussions. We acknowledge financial support from NASA grant ATP12-ATP12-0017, as well as computational resources from NSF XSEDE and the University of Michigan. Simulations described in this work were performed using the publicly-available Enzo code \citep{Enzo}, which is the product of a collaborative effort of many independent scientists from numerous institutions around the world. The simulation data is analyzed using the publicly available yt visualization package \citep{yt}. We are grateful to the yt development team and the yt community for their support.

\end{document}